\documentclass[conference]{IEEEtran}
\usepackage{balance}
\IEEEoverridecommandlockouts
\usepackage{xcolor}
\usepackage{xspace}
\usepackage{subcaption}
\usepackage{pifont}
\usepackage{times}
\usepackage{graphbox}
\usepackage{graphicx}
\usepackage{epstopdf}
\usepackage{dblfloatfix} % To enable figures at the bottom of page
\usepackage[htt]{hyphenat}
\usepackage{fancyhdr}
\usepackage{setspace}
\usepackage{relsize}
\usepackage{textcomp}
\usepackage{circledsteps}
\usepackage{url}
\usepackage[super]{nth}
\usepackage{comment}
\usepackage{transparent}
\usepackage{wrapfig}
\usepackage{rotating}

\usepackage{CJK}
\usepackage[utf8]{inputenc}
\usepackage[T1]{fontenc}
\usepackage{soul}
\usepackage{xcolor}
\usepackage{kotex}
\usepackage{lipsum}
\usepackage[normalem]{ulem}
\makeatletter
\def\SOUL@hlpreamble{%
\setul{\dimexpr\dp\strutbox-2pt}{\dimexpr\ht\strutbox+\dp\strutbox-2pt\relax}
\let\SOUL@stcolor\SOUL@hlcolor
\SOUL@stpreamble
}
\makeatother

\newcommand\khlc[1][yellow]{
  \bgroup
  \markoverwith{\textcolor{#1}{\rule[-.5ex]{1pt}{2.5ex}}}
  \ULon
}

\newcommand{\todo}[1]{\bgroup\color{white}\textbf{\khlc[black]{TODO: [#1]}}\egroup\xspace}
\newcommand{\fixme}[1]{\bgroup\color{red}\textbf{\khlc{FIXME: [#1]}}\egroup\xspace}
\newcommand{\pointer}[1]{\bgroup\color{white}\textbf{\khlc[red]{POINTER: [#1 is working here]}}\egroup\xspace}
\newcommand{\reviewer}[1]{\bgroup\color{blue}#1\egroup\xspace}
\newcommand{\ranswer}[1]{\bgroup\color{red}#1\egroup\xspace}
\usepackage{enumitem}
\usepackage{amssymb}
\usepackage[framemethod=tikz]{mdframed}
\usepackage[numbers,sort&compress]{natbib}
\tikzset{
    vertical align/.style={
        baseline=-.5*(height("$+$")-depth("$+$"))
    }
}

\usepackage{calc}
\usepackage{amsmath,algorithm,algpseudocode}
\usepackage{mathtools}
\usepackage{cases}
\algrenewcommand\algorithmicindent{0.5em}
\makeatletter
\algrenewcommand\ALG@beginalgorithmic{\footnotesize}
\makeatother
\makeatletter
\renewcommand{\Function}[2]{%
  \csname ALG@cmd@\ALG@L @Function\endcsname{#1}{#2}%
  \def\jayden@currentfunction{#1}%
}
\newcommand{\funclabel}[1]{%
  \@bsphack
  \protected@write\@auxout{}{%
    \string\newlabel{#1}{{\jayden@currentfunction}{\thepage}}%
  }%
  \@esphack
}
\makeatother

\usepackage{tabularx}
\usepackage{array}
\usepackage{multirow}
\usepackage{booktabs}
\usepackage[para]{threeparttable}
\usepackage{colortbl}
\usepackage{multirow}

\makeatletter
\def\hlinewd#1{%
\noalign{\ifnum0=`}\fi\hrule \@height #1 %
\futurelet\reserved@a\@xhline}
\makeatother
\usepackage{hhline}
\newcolumntype{C}{>{\centering\arraybackslash}X}
\usepackage[framemethod=tikz]{mdframed}
\usepackage{marginnote}
\usepackage{pdfcomment}
\mdfsetup{
  hidealllines=true,
  innerleftmargin=2pt,
  innerrightmargin=2pt,
  innertopmargin=2pt,
  innerbottommargin=2pt,
  leftmargin=-2pt,
  rightmargin=-2pt,
  skipabove=-2pt,
  skipbelow=-2pt,
  backgroundcolor=blue!20}

\newlength{\markerHeight}
\newlength{\markerMargin}
\newlength{\linespace}
\newlength{\linedepth}

\sbox0{yf}
\definecolor{mylime}{RGB}{205, 220, 57}
\definecolor{mygreen}{RGB}{60, 200, 0}
\definecolor{myblue}{RGB}{0, 51, 204}

\colorlet{soulred}{red!20}
\colorlet{soulgreen}{green!20}
\colorlet{soulblue}{blue!20}

\usepackage{enumitem}
\setlistdepth{5}
\renewlist{itemize}{itemize}{5}
\setlist[itemize,1]{label=$\bullet$}
\setlist[itemize,2]{label=$\circ$}
\setlist[itemize,3]{label=$\ast$}
\setlist[itemize,4]{label=-}
\setlist[itemize,5]{label=$\cdot$}

\makeatletter
\def\SOUL@hlpreamble{%
\setul{\dimexpr\dp\strutbox-2pt}{\dimexpr\ht\strutbox+\dp\strutbox-2pt\relax}
\let\SOUL@stcolor\SOUL@hlcolor
\SOUL@stpreamble
}
\makeatother

\usepackage{hyperref}
\hypersetup{
    hidelinks
}

\usepackage{amsmath,amssymb,amsfonts}
\usepackage{graphicx}
\usepackage{textcomp}
\usepackage{xcolor}
\immediate\write18{refine_ref.sh}

\begin{document}

\title{MPI-over-CXL: Enhancing Communication Efficiency in Distributed HPC Systems}

\author{
\IEEEauthorblockN{
    Miryeong Kwon, 
    Donghyun Gouk, 
    Hyein Woo, 
    Junhee Kim, 
    Jinwoo Baek, 
    Kyungkuk Nam,
    Sangyoon Ji, \\
    Jiseon Kim, 
    Hanyeoreum Bae, 
    Junhyeok Jang, 
    Hyunwoo You, 
    Junseok Moon, 
    Myoungsoo Jung}
\IEEEauthorblockA{Panmnesia, Inc. \\
\href{https://panmnesia.com/}{https://panmnesia.com}}
}

\maketitle

\begin{abstract}
MPI implementations commonly rely on explicit memory-copy operations, incurring overhead from redundant data movement and buffer management. This overhead notably impacts HPC workloads involving intensive inter-processor communication. In response, we introduce MPI-over-CXL, a novel MPI communication paradigm leveraging CXL, which provides cache-coherent shared memory across multiple hosts.

MPI-over-CXL replaces traditional data-copy methods with direct shared memory access, significantly reducing communication latency and memory bandwidth usage. By mapping shared memory regions directly into the virtual address spaces of MPI processes, our design enables efficient pointer-based communication, eliminating redundant copying operations. To validate this approach, we implement a comprehensive hardware and software environment, including a custom CXL 3.2 controller, FPGA-based multi-host emulation, and dedicated software stack.

Our evaluations using representative benchmarks demonstrate substantial performance improvements over conventional MPI systems, underscoring MPI-over-CXL's potential to enhance efficiency and scalability in large-scale HPC environments.
\end{abstract}

\section{Introduction}
Message passing interface (MPI) is a widely adopted communication protocol facilitating efficient data exchange among processors in distributed computing systems \cite{Padua2011,mpi1,mpi2}. MPI is particularly prevalent in applications demanding frequent inter-processor communication, such as parallel processing and high-performance computing (HPC) workloads \cite{vay2018warp,spalart2016role,Graph500,NASA.IS,lauguna2019large,mpiapp,haloapp}. For instance, HPC simulations (e.g., computational fluid dynamics and plasma simulations) require extensive computational resources, involving multiple processors collaboratively executing tasks and regularly exchanging data. Such data exchanges significantly influence the overall system performance, accounting for a substantial portion of total execution time.

Traditional MPI implementations typically rely on memory-copy mechanisms, as each processor maintains its own independent memory address space. This approach results in high memory access latency and inefficient memory utilization due to redundant data storage. Furthermore, most existing MPI solutions utilize remote direct memory access (RDMA), which involves complex initialization processes and high communication overhead \cite{zambre2019breaking,mpioverhead,mpioverhead2,mpioverhead4}, ultimately limiting system performance.

Recently introduced \emph{Compute Express Link} (CXL) \cite{cxl3} has emerged as a promising next-generation high-performance interconnect technology. CXL provides hardware-based cache coherency between processors and memory devices, enabling efficient memory sharing among multiple hosts \cite{memsharing,memsharing1,memsharing2}. Leveraging CXL's memory-sharing capabilities, processors can exchange data without memory copying, simply by transferring pointers, thereby significantly reducing data access latency and redundant memory usage \cite{huang2025txcocket,zhang2023partial}. In addition, as CXL employs straightforward memory semantics, it minimizes communication overhead by eliminating the complex initialization processes and software interventions inherent to RDMA.
\begin{figure}[t]
    \centering
    \includegraphics[width=\linewidth]{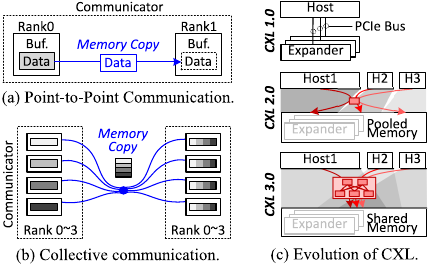}
    \vspace{-15pt}
    \caption{Traditional MPI and emerging CXL.} \label{fig:background}
    \vspace{-15pt}
\end{figure}
\subsection{MPI Data Movement Patterns in Parallel Computing}

In this work, we propose \textit{MPI-over-CXL}, a novel paradigm that transforms traditional memory-copy-based MPI into memory-sharing-based MPI using CXL technology. MPI-over-CXL maps shared memory spaces onto identical virtual addresses across all processors, enabling simple pointer exchanges rather than actual data transmission. By relocating MPI's message queues from local processor memory to shared memory and exchanging pointers referencing data instead of transferring data itself \cite{shmem,openshmem1,sur2006rdma}, we substantially reduce memory access latency and data copying overhead, thereby streamlining the communication process. Especially, for irregular memory access patterns of sparse workloads, RDMA-based MPI inevitably requires data reformatting \cite{wu2004high,traff2023uniform,gainaru2016using}, whereas MPI-over-CXL’s pointer-based fine-grained communication enables direct access to the native data structure.

We implemented and validated MPI-over-CXL using multiple CXL 3.2 controllers whose silicon has been proven using a 4nm advanced process technology. For system-level evaluations, we constructed an FPGA-based hardware prototype incorporating all essential switches and endpoints necessary to form a complete CXL interconnect network. Evaluation results using various MPI benchmarks demonstrate that MPI-over-CXL significantly enhances performance, achieving up to 1.6$\times$ faster overall execution and, on average, 8.4$\times$ faster communication compared to traditional memory-copy-based MPI implementations. Moreover, MPI-over-CXL promises even greater performance benefits in larger-scale cluster environments with increased processor counts.

\section{Background}
In HPC systems, applications frequently surpass the computational and memory capabilities of individual processors, necessitating parallel execution across multiple processors. To enable efficient inter-processor data exchange, MPI is widely adopted, offering standardized communication APIs \cite{mpi2,zambre2019breaking}. Despite its broad use, traditional MPI implementations incur significant overhead due to reliance on memory-copy operations, leading to increased memory access latency and redundant data movement \cite{gainaru2016using,mpioverhead,mpioverhead2,mpioverhead4}.

MPI-based scientific applications \cite{keppens2023mpi,ly2023alisim} often exhibit a combination of two communication methods to share data among processors: point-to-point and collective communication \cite{memorycopyredun,gainaru2016using,mpioverhead4,chen2023mpi} (cf. Figures \ref{fig:background}a and \ref{fig:background}b). In MPI, each process belongs to a communicator and is identified by a unique rank within it. Point-to-point communication involves data exchange between two processes, where one process issues a send operation and the other performs a matching receive based on their ranks. Collective communication, on the other hand, coordinates data movement among all processes in the communicator \cite{gainaru2016using,mpi5,multiplenode}.

Note that these MPI communication methods introduce significant synchronization overhead, exacerbated by traditional memory-copy-based MPI implementations \cite{balaji2010,bierbaum2022towards}. These overheads intensify as processors increase \cite{mpinodecomlexity,mpimanyprocessor,hofinger2017modelling,multiplenode}, highlighting the need for alternative communication paradigms, such as MPI-over-CXL, to effectively mitigate these limitations.

\vspace{-5pt}
\subsection{Cache-Coherent Interconnect for Multi-Host Environments}

CXL has emerged as a cache-coherent interconnect designed to overcome communication and memory-sharing limitations inherent in traditional heterogeneous computing environments. Built upon \emph{PCI Express} (PCIe) \cite{pcie}, CXL provides standardized memory semantics supporting coherent communication between multiple hosts and diverse peripheral devices, including accelerators and memory expanders \cite{gouk2022direct,cxl1,cxl2,cxl3}. As workloads and data volumes scale, CXL facilitates resource sharing across nodes, enabling effective utilization of memory resources and streamlined inter-device communication.

\noindent \textbf{Protocol components.}
CXL primarily consists of three protocols: CXL.io, CXL.cache, and CXL.mem. Specifically, CXL.io, derived from PCIe, manages device configuration, interrupts, and standard I/O operations using PCIe transaction layer packets. CXL.cache facilitates coherent caching among devices and hosts by supporting coherence transactions, such as snoop requests and responses, to maintain cache consistency. CXL.mem enables devices to directly perform coherent reads and writes to shared memory, which is crucial for memory pooling and shared memory configurations.

\noindent \textbf{CXL progression.}
CXL has evolved from version 1.0 to 3.0 \cite{cxl1,cxl2,cxl3}, introducing new capabilities for different computing scenarios as shown in Figure \ref{fig:background}c. Initially, CXL 1.0 supported cache-coherent memory expansion for single-host environments. CXL 2.0 extended this by enabling memory pooling, allowing multiple hosts to dynamically allocate dedicated memory segments from shared memory resources. The latest version, CXL 3.0, further extends memory sharing by supporting simultaneous shared memory access by multiple hosts, thus improving memory utilization and enabling more efficient resource sharing.

\noindent \textbf{Implementation complexities.}
However, implementing CXL 3.0 introduces several technical challenges. Multi-path responses in multi-level switch topologies complicate routing responses back to originating hosts. In addition, coherence management across multiple hosts inherently adds latency, as maintaining cache coherence typically involves multiple communication round-trips among hosts, switches, and memory expanders. Moreover, standardized methods for managing device coherent memory regions, which ensure coherent shared access, remain undefined, complicating implementations.

\noindent \textbf{Implementation complexities of shared memory.} In a typical CXL configuration, multiple hosts access shared memory resources through a CXL switch. Hosts dynamically allocate memory from multiple memory expanders connected via switch ports. CXL 3.0 also supports concurrent access to shared memory regions by multiple hosts within the same memory expander. Managing shared access requires additional coherence protocols and metadata mechanisms to maintain data consistency and correctness. Addressing coherence management and multi-host memory access underscores the importance of solutions such as MPI-over-CXL.

\section{Design of Scalable MPI Communication}
\begin{figure}
    \centering
    \includegraphics[width=\linewidth]{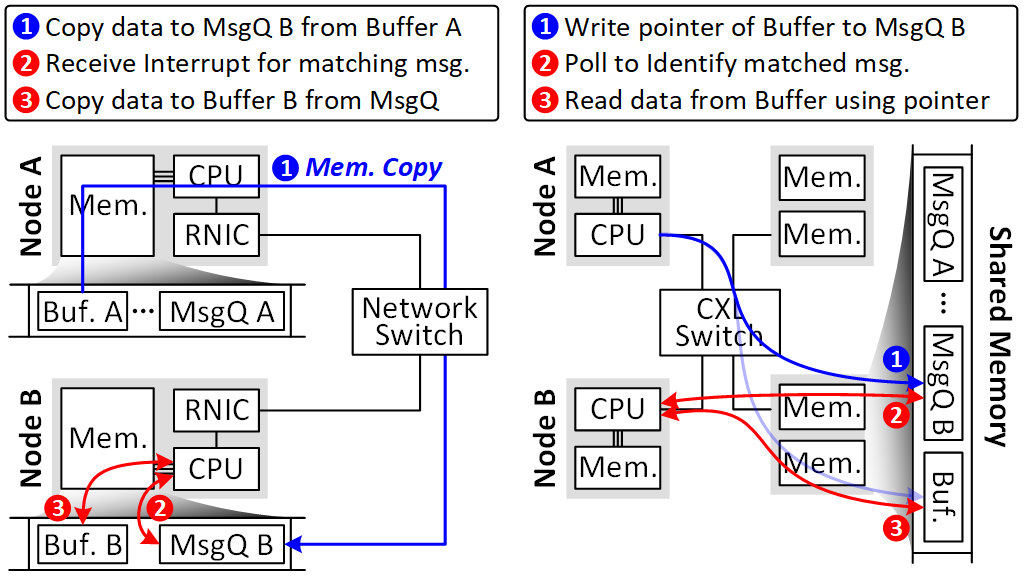}
    \begin{subfigure}{\linewidth}
        \begin{tabularx}{\textwidth}{
            p{\dimexpr.49\linewidth-2\tabcolsep-1.3333\arrayrulewidth}
            p{\dimexpr.49\linewidth-2\tabcolsep-1.3333\arrayrulewidth}
            }
            \caption{Traditional paradigm.} \label{fig:overview1} &
            \caption{MPI-over-CXL.} \label{fig:overview2}
        \end{tabularx}
    \end{subfigure}
    \vspace{-14pt}
    \caption{Comparison of MPI communication paradigms.} \label{fig:overview}
    \vspace{-12pt}
\end{figure}

Traditional MPI implementations use explicit memory-copy operations, where processes copy data between regions. This introduces significant overhead due to redundant data movements, higher memory bandwidth consumption, and buffer management complexity \cite{memorycopyredun,mpioverhead,mpioverhead2,mpioverhead4}. To mitigate these issues, we propose \textit{MPI-over-CXL}, a novel MPI communication method using the shared memory features provided by CXL. It leverages CXL's coherent shared memory capabilities, allowing multiple processors direct access to a unified memory space without redundant copying. Specifically, our MPI-over-CXL design
replaces traditional memory-copy methods with direct shared memory access by leveraging CXL 3.0 supporting coherent memory sharing among hosts \cite{memsharing,memsharing1,memsharing2}. In this design, each MPI process maps shared memory directly into its virtual address space, enabling pointer-based communication without explicit data transfers. Thus, sending processes update shared memory content, while receiving processes directly access updated data, removing intermediate buffer handling.

Compared to standard MPI methods, our MPI-over-CXL approach reduces communication latency and memory bandwidth usage by eliminating redundant memory operations. It also enhances memory efficiency by allowing concurrent access to shared physical memory.

\subsection{CXL-Enabled MPI Message Handling}
As shown in Figure \ref{fig:overview}, MPI-over-CXL transforms traditional MPI communication by removing explicit data-copy operations, using cache-coherent shared memory features provided by CXL. It enables MPI processes to directly interact with unified memory regions through pointer-based methods, simplifying communication. As each MPI process maps shared memory allocated from CXL memory expanders into its virtual address space, traditional message queues, previously local, can now be maintained in shared memory regions. MPI ranks handle communication by directly accessing these shared queues, thereby removing redundant memory copies.

To implement this communication model, MPI-over-CXL specifically comprises two key architectural components. First, shared message queues allow direct exchanges between MPI ranks within coherent memory, enabling interactions through pointer operations without intermediate buffers. Second, shared memory management uses CXL memory expanders to allocate and maintain shared buffers, providing direct access through pointers to lower latency and reduce memory bandwidth usage. During operation, sender processes directly write data into shared queues, while receiver processes poll these queues to directly retrieve data.This guarantees that exchanged pointers always refer to data in the shared memory region rather than the sender’s local buffer, allowing both sender and receiver to reuse or modify their own local buffers independently at any given time. Synchronization primitives, such as barriers, ensure data consistency and avoid race conditions, maintaining correctness. By leveraging shared-memory-based communication \cite{openshmem1}, MPI-over-CXL significantly reduces latency and memory overhead compared to standard MPI methods. These advantages become more pronounced with increased processor counts, enhancing scalability and HPC system performance.

\begin{figure}
    \centering
    \includegraphics[width=\linewidth]{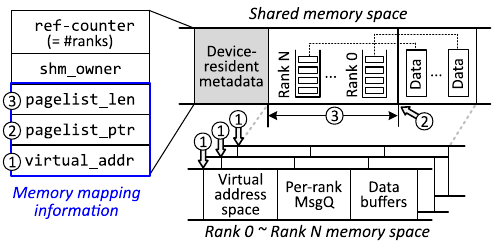}
    \begin{subfigure}{\linewidth}
        \begin{tabularx}{\textwidth}{
            p{\dimexpr.30\linewidth-2\tabcolsep-1.3333\arrayrulewidth}
            p{\dimexpr.49\linewidth-2\tabcolsep-1.3333\arrayrulewidth}
            }
            \caption{Metadata.} \label{fig:metadata} &
            \caption{Shared memory regions.} \label{fig:sharedmem}
        \end{tabularx}
    \end{subfigure}
    \vspace{-16pt}
    \caption{Memory management and data structures.} \label{fig:memory_mgmt}
    \vspace{-12pt}
\end{figure}

\subsection{Memory Management and Data Structures}
\noindent \textbf{Memory handling.}
Effective memory management and consistent address mapping are essential for MPI-over-CXL. To this end, each MPI rank maps shared memory regions, allocated from memory expanders, into its virtual address space (cf. Figure \ref{fig:memory_mgmt}). This unified memory approach allows MPI ranks to perform direct inter-process communication by reading from and writing to shared buffers without redundant data copying. Specifically, MPI-over-CXL ensures shared memory regions are consistently mapped to identical virtual addresses across all participating processes. When an MPI rank accesses a shared memory region for the first time, the operating system resolves page faults by directly allocating physical memory pages from the CXL memory expander, thereby ensuring consistent and efficient memory access across processes.

\begin{figure}
    \centering
    \includegraphics[width=\linewidth]{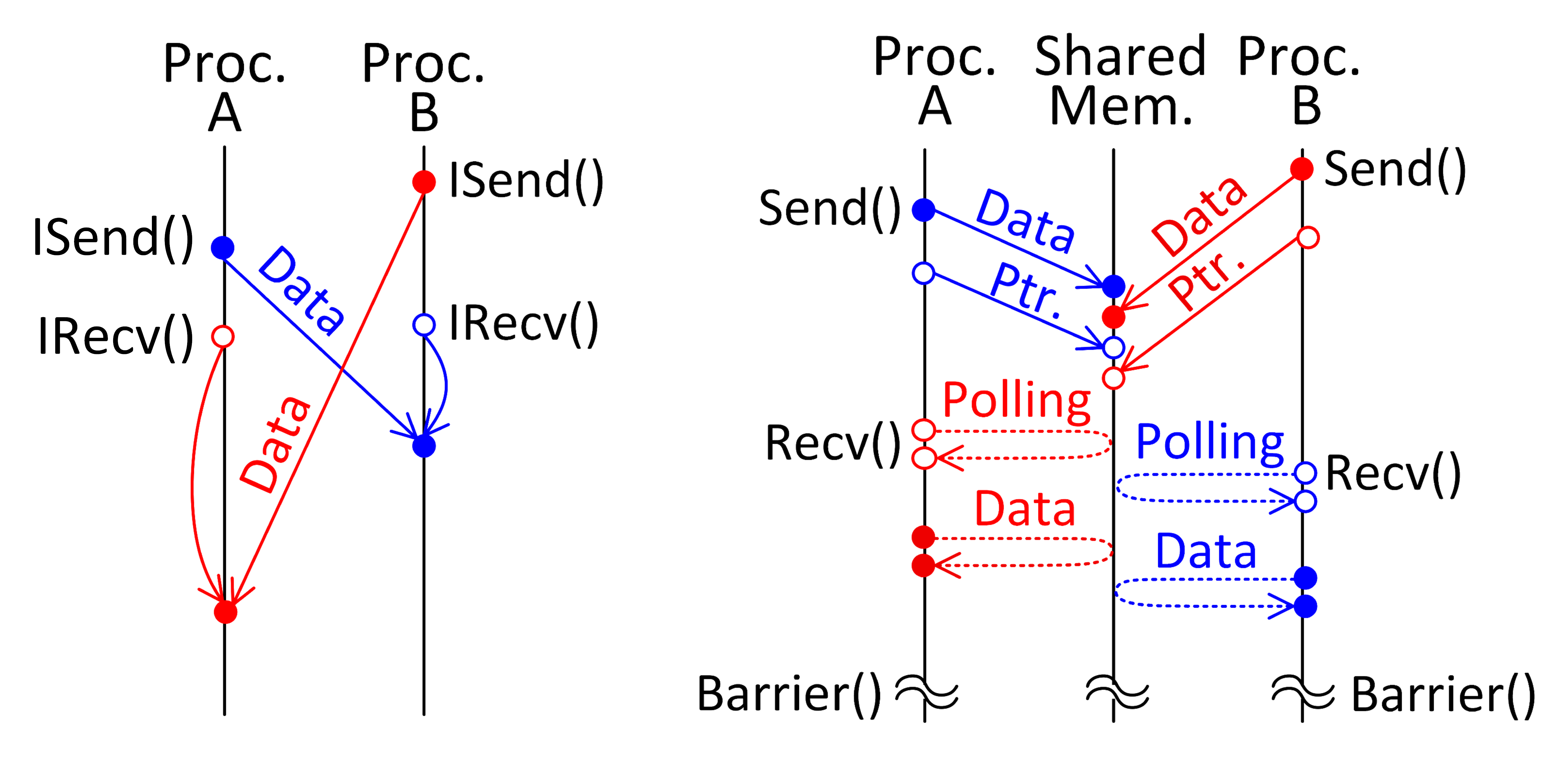}
    \begin{subfigure}{\linewidth}
        \begin{tabularx}{\textwidth}{
            p{\dimexpr.46\linewidth-2\tabcolsep-1.3333\arrayrulewidth}
            p{\dimexpr.56\linewidth-2\tabcolsep-1.3333\arrayrulewidth}
            }
            \caption{Traditional paradigm.} \label{fig:implementation1} &
            \caption{MPI-over-CXL.} \label{fig:implementation2}
        \end{tabularx}
    \end{subfigure}
    \vspace{-16pt}
    \caption{Implementation of MPI asynchronous operations.} \label{fig:implementation}
    \vspace{-12pt}
\end{figure}

\noindent \textbf{Queue and data structure.}
Traditional MPI message queues, previously in local memory, are relocated into these shared memory regions. Each MPI rank manages its dedicated message queue within the unified memory space, facilitating efficient pointer-based access. Communication occurs by exchanging pointers referencing data within the shared memory region. Specifically, instead of copying data, a sender places the data into shared memory and simply inserts the corresponding pointer into the receiver's message queue. Compared to traditional MPI methods involving repeated data copying (e.g. explicit data packing/unpacking \cite{santhanaraman2005designing,hashmi2020falcon,gainaru2016using}, data transfer between local memory and send/receive buffers \cite{ma2021thinking,kong2023understanding}), this approach significantly reduces communication latency and memory bandwidth usage \cite{opt1,opt2}. Device-resident metadata stored in the CXL memory expander tracks characteristics of shared memory regions, such as region identifiers, page mappings, and reference counts. This metadata maintains accurate state information and ensures conflict-free and coherent memory interactions among multiple MPI ranks. Additionally, small-sized data can be placed directly in a dedicated buffer space in the shared memory without a pointer exchange, thereby simplifying the transmission procedure in a manner similar to the MPI eager mode while still preserving the benefit of eliminating copy overhead in MPI-over-CXL.

MPI-over-CXL integrates tightly-coupled hardware and software components, each addressing critical aspects of shared memory management and MPI operations. By focusing on unified memory management, consistent address mapping, and direct pointer-based communication, MPI-over-CXL provides a scalable solution specifically tailored for high-performance computing workloads.

\subsection{Communication Procedure and Synchronization}

MPI-over-CXL manages communication through shared message queues located within a unified memory region accessible by all participating MPI ranks. Each rank maintains its dedicated message queue, enabling efficient concurrent data exchanges and removing the need for intermediate buffers.

Figure \ref{fig:implementation} shows the communication procedure among different processors through MPI-over-CXL. Communication proceeds as follows: sender processes directly place data into shared memory and insert pointers referencing this data into the receiver’s message queue. The receiver continuously monitors its message queue entries and accesses the data directly upon detecting a valid pointer, avoiding additional copying steps. Message queue entries include pointers to the shared memory data and associated status flags indicating data availability, facilitating efficient management and reuse. After the receiver accesses and consumes the data, it updates the status flags, allowing efficient reuse of message queue entries.

To ensure data consistency and avoid race conditions, MPI-over-CXL incorporates fine-grained synchronization mechanisms such as atomic operations and locks within message queue structures. Lock acquisition on metadata ensures that only one process can write to a given shared memory region at a time, and MPI barriers enforce synchronization at defined intervals, maintaining in-order send/receive execution, thereby preventing buffer pollution scenarios where a sender overwrites data before a receiver consumes it or a receiver reads before data is fully written. In addition, MPI-over-CXL relies on CXL's inherent cache coherence protocols to ensure data visibility across processors.

\noindent \textbf{Dynamic polling adjustment.}
MPI-over-CXL also implements optimized polling strategies, adjusting polling frequencies dynamically based on communication traffic, reducing unnecessary CPU usage during idle periods. Event-driven notifications, such as interrupts triggered by hardware upon new message availability, effectively complement polling by significantly reducing idle CPU cycles. Continuous polling introduces little overhead in this context because lock acquisition only involves accessing a shared memory region under hardware-maintained cache coherence. Message queues are cached after each process's first access, and subsequent reads can be served locally. Whenever other processes modify the message queue, hardware-triggered cache invalidations ensure coherence, thereby substantially reducing both latency and communication overhead compared to traditional polling mechanisms.

Moreover, MPI-over-CXL addresses challenges related to asynchronous operations (e.g., \texttt{MPI\_Isend}, \texttt{MPI\_Irecv}), such as Read-After-Write hazards and race conditions. It introduces lightweight controlled synchronization points to ensure data and message visibility, enabling the safe progress of asynchronous MPI operations. For instance, an \texttt{MPI\_Isend} operation is internally wrapped as synchronous, where the sender explicitly waits at a barrier to confirm message delivery completion, ensuring correct message ordering and eliminating potential race conditions. Unlike traditional distributed-memory MPI where a blocking send cannot complete until a matching receive is posted, MPI-over-CXL completes a send once data is written into the shared memory and a pointer with status flags is enqueued into the receiver’s queue. Because this does not depend on the receiver’s immediate action and CXL coherence guarantees visibility, the classical deadlock condition does not occur. Despite introducing additional synchronization, this strategy minimally impacts performance due to significantly reduced communication overhead achieved by eliminating redundant buffer copying.

\section{Prototype Implementation and Evaluation}
\begin{figure}
    \centering
    \begin{subfigure} {0.41\linewidth}
        \includegraphics[height=3.6cm]{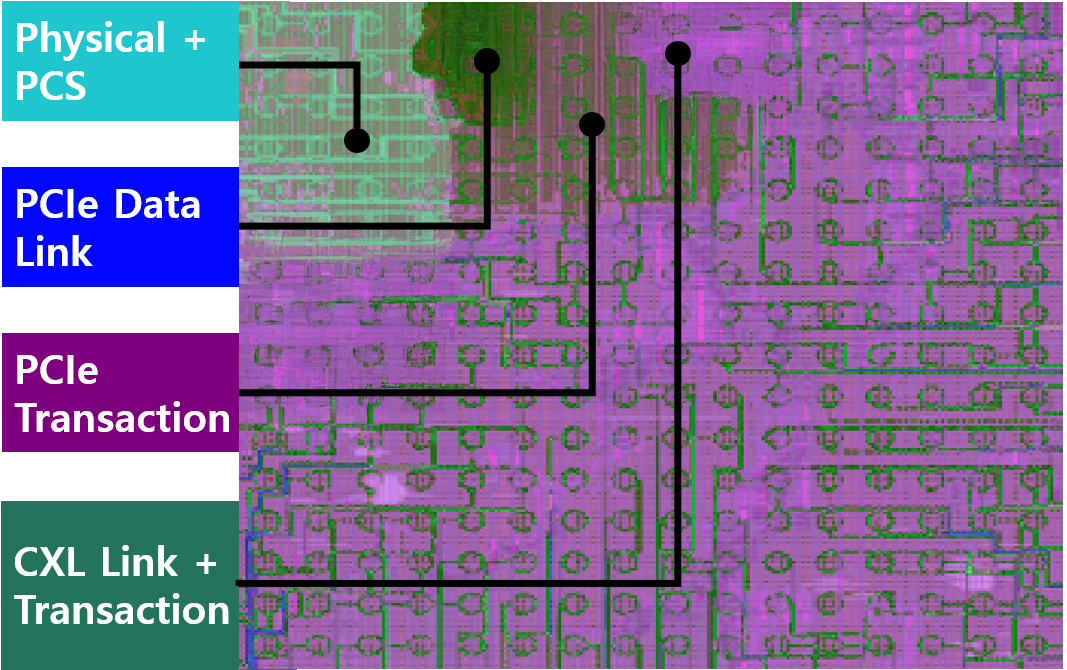}
    \end{subfigure}
    \hfill
    \begin{subfigure} {0.57\linewidth}
        \includegraphics[height=3.6cm]{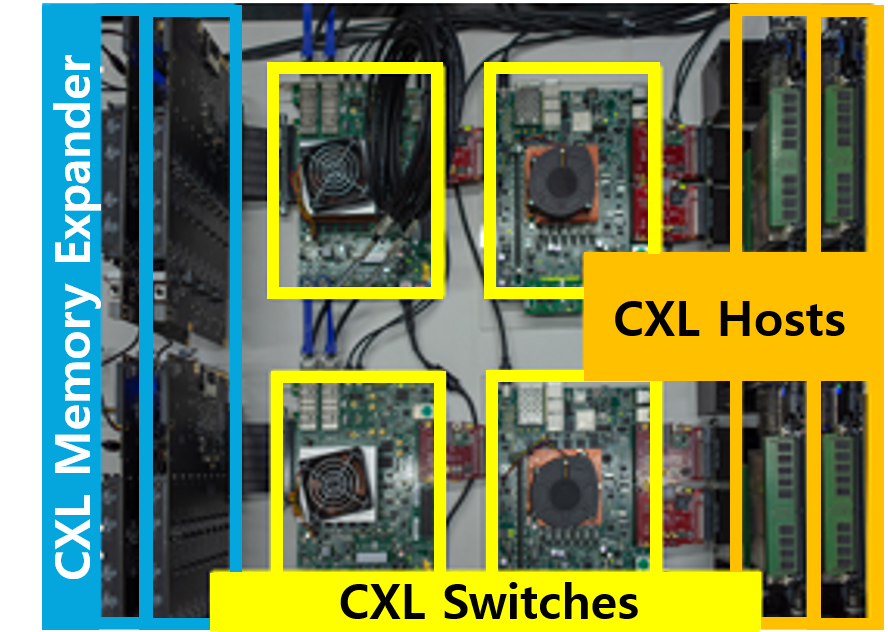}
    \end{subfigure}
    \begin{subfigure}{\linewidth}
        \begin{tabularx}{\textwidth}{
            p{\dimexpr.41\linewidth-2\tabcolsep-1.3333\arrayrulewidth}
            p{\dimexpr.57\linewidth-2\tabcolsep-1.3333\arrayrulewidth}
            }
            \caption{CXL 3.2 controller.} \label{fig:controller} &
            \captionsetup{margin={5mm,0pt}}
            \caption{RTL-emulated environment.} \label{fig:prototype}
        \end{tabularx}
    \end{subfigure}
    \vspace{-16pt}
    \caption{Evaluation environment.} \label{fig:overall}
    \vspace{-12pt}
\end{figure}

\noindent \textbf{CXL controller and system emulation.}
We design and implement a CXL 3.2 controller, which integrates PCIe 6.0 \cite{pcie} physical media attachment, a physical coding sublayer specific to CXL, a pipe interface, and complete hardware stacks including link and transaction layers (cf. Figure \ref{fig:controller}). This design was validated using a 4nm semiconductor process. For evaluating MPI-over-CXL at the system-level, we developed an RTL-emulated environment consisting of four independent hosts interconnected with four memory expanders via four FPGA-based CXL switches as shown in Figure \ref{fig:prototype}. Each host includes custom dual-core RISC-V processors, fully supporting CXL 3.x and integrating root port (RP) features. FPGA-based memory expanders provide 1TB of coherent shared memory using eight DDR4 DRAM modules, enabling realistic testing of memory-intensive workloads. Multi-port CXL switches interconnect via fabric links, validating scalability and architectural flexibility.

\begin{figure} [!t]
    \centering
    \begin{subfigure} {0.49\linewidth}
        \includegraphics[height=3.6cm]{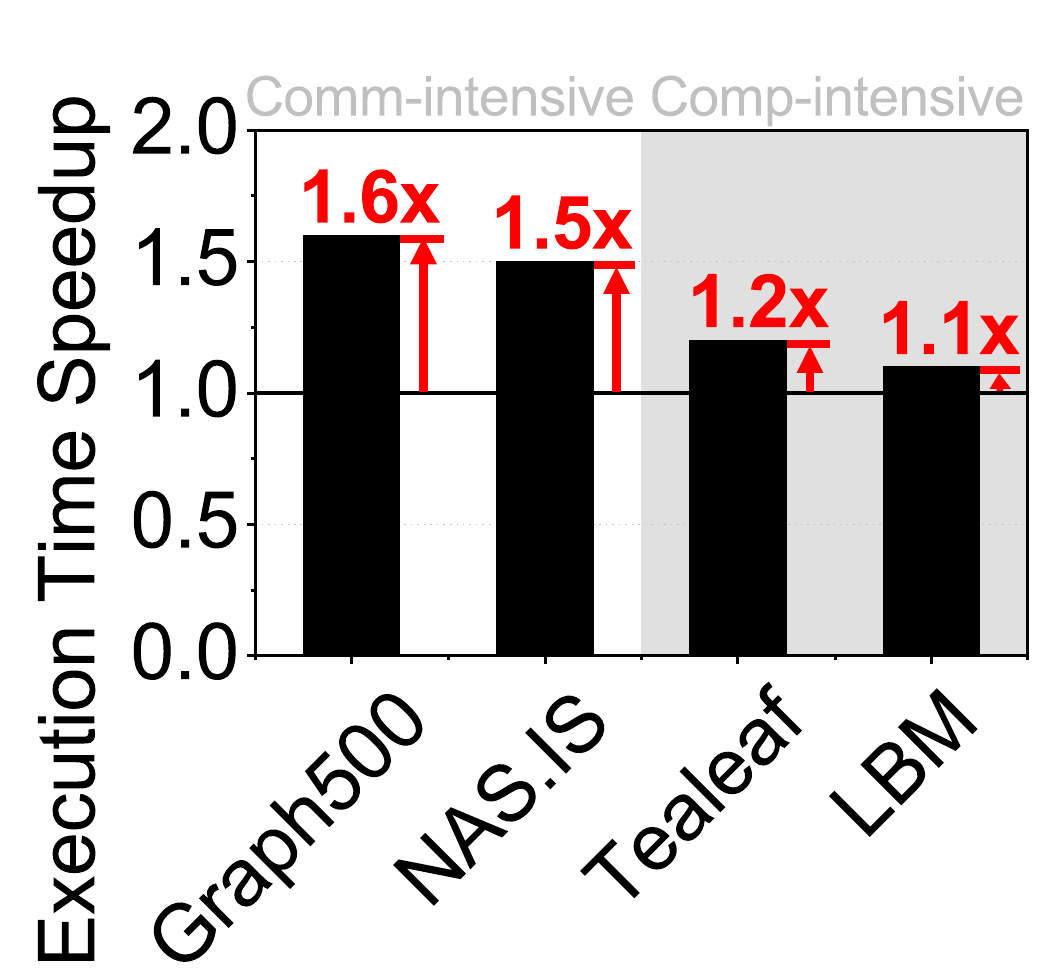}
    \end{subfigure}
    \hfill
    \begin{subfigure} {0.49\linewidth}
        \includegraphics[height=3.6cm]{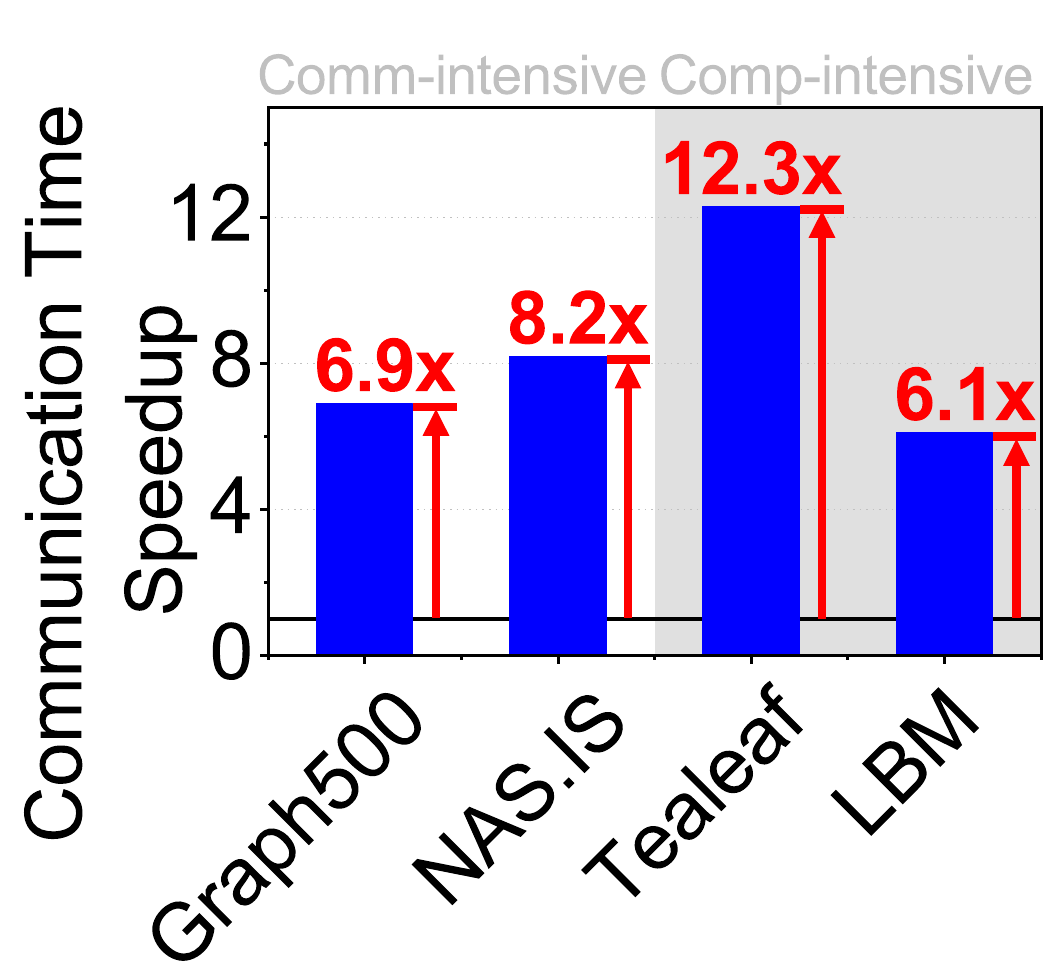}
    \end{subfigure}
    \begin{subfigure}{\linewidth}
        \begin{tabularx}{\textwidth}{
            p{\dimexpr.52\linewidth-2\tabcolsep-1.3333\arrayrulewidth}
            p{\dimexpr.49\linewidth-2\tabcolsep-1.3333\arrayrulewidth}
            }
            \caption{Execution Speedup.} \label{fig:exespeedup} &
            \caption{Communication Speedup.} \label{fig:comspeedup}
        \end{tabularx}
    \end{subfigure}
    \caption{Speedup.} \label{fig:speedup}
\end{figure}

\noindent \textbf{Software stack and validation.}
Our software includes a custom kernel driver ($\sim$2,352 LoC) in RISC-V Linux for managing shared memory, virtual addressing, and synchronization. A dedicated user-space MPI-over-CXL library seamlessly integrates with standard MPI applications, with minimal adaptation to existing MPI calls. Validation involved rigorous tests on commercial FPGA boards, confirming the correct operation of host-memory coherence, shared memory management, and switch functionality, accurately representing real-world CXL environments.

\noindent \textbf{Workloads.}
To comprehensively evaluate the performance of MPI-over-CXL, we selected a set of representative HPC workloads characterized by distinct communication and computation profiles. While all workloads incorporate both point-to-point and collective MPI operations, the relative contribution of communication varies across applications. The Graph500 benchmark \cite{Graph500}, which performs distributed breadth-first search, is primarily communication-bound and dominated by point-to-point interactions. NAS Integer Sort (NAS.IS) \cite{NASA.IS} imposes a high collective communication burden due to its global data redistribution phase. On the other hand, the Tealeaf application \cite{tealeaf}, modeling linear heat conduction, and the Lattice-Boltzmann Method (LBM) benchmark \cite{LBM}, representing computational fluid dynamics, are both computation-intensive; communication overhead in these cases constitutes a minor fraction of the overall execution time. Taken together, these benchmarks capture a broad spectrum of communication behaviors in scientific computing and effectively illustrate the performance gains achievable with MPI-over-CXL.

\noindent \textbf{Preliminary evaluation.}
We evaluate the MPI-over-CXL prototype, assessing both functional correctness and performance improvements compared to a baseline Distributed Shared Memory (DSM) system \cite{cai2018efficient} using CXL 2.0 without cache coherence. By comparing against a CXL 2.0–based baseline, we validate the effectiveness of our key idea, leveraging shared memory to eliminate the overhead of memory copying and RDMA-based communication. We evaluate the effectiveness of our shared-memory-based solution by analyzing its ability to reduce RDMA communication overhead compared to the CXL 2.0 baseline. Our evaluations utilize advanced interconnect technologies, specifically PCIe Gen 6 and ConnectX-8, to reflect realistic computing environments.
The evaluation consists of two primary analyses. First, we measured the normalized total execution time and communication time to quantify overall speedup and isolate MPI-related communication overhead from computation time. Second, we conducted a strong scaling test to examine how performance scales with increasing processor counts under fixed problem sizes.

As shown in Figure \ref{fig:speedup}, the normalized execution-time results highlight the performance advantages of MPI-over-CXL over the baseline DSM system. Specifically, MPI-over-CXL achieved speedups ranging from 1.1$\times$ to 1.6$\times$. Communication overhead was also significantly reduced, averaging 8.4$\times$ across benchmarks. This improvement stems from MPI-over-CXL's elimination of redundant memory-copy operations in traditional MPI methods. By using coherent shared memory through direct pointer-based interactions, MPI-over-CXL significantly decreases the communication overhead. In terms of pure communication time, tealeaf achieves the largest speedup of 12.3$\times$ over memory-copy–based MPI, as our approach effectively reduces the communication overhead \cite{spechpc} that is particularly high in tealeaf among the SpecHPC workloads. From an end-to-end execution time perspective, communication-intensive workloads (i.e., Graph500, NAS.IS) show relatively lower pure communication speedups; however, since communication accounts for nearly half of their overall runtime, they still yield substantial average speedups.

\begin{figure}[!t]
    \centering
    \includegraphics[width=1.105\linewidth]{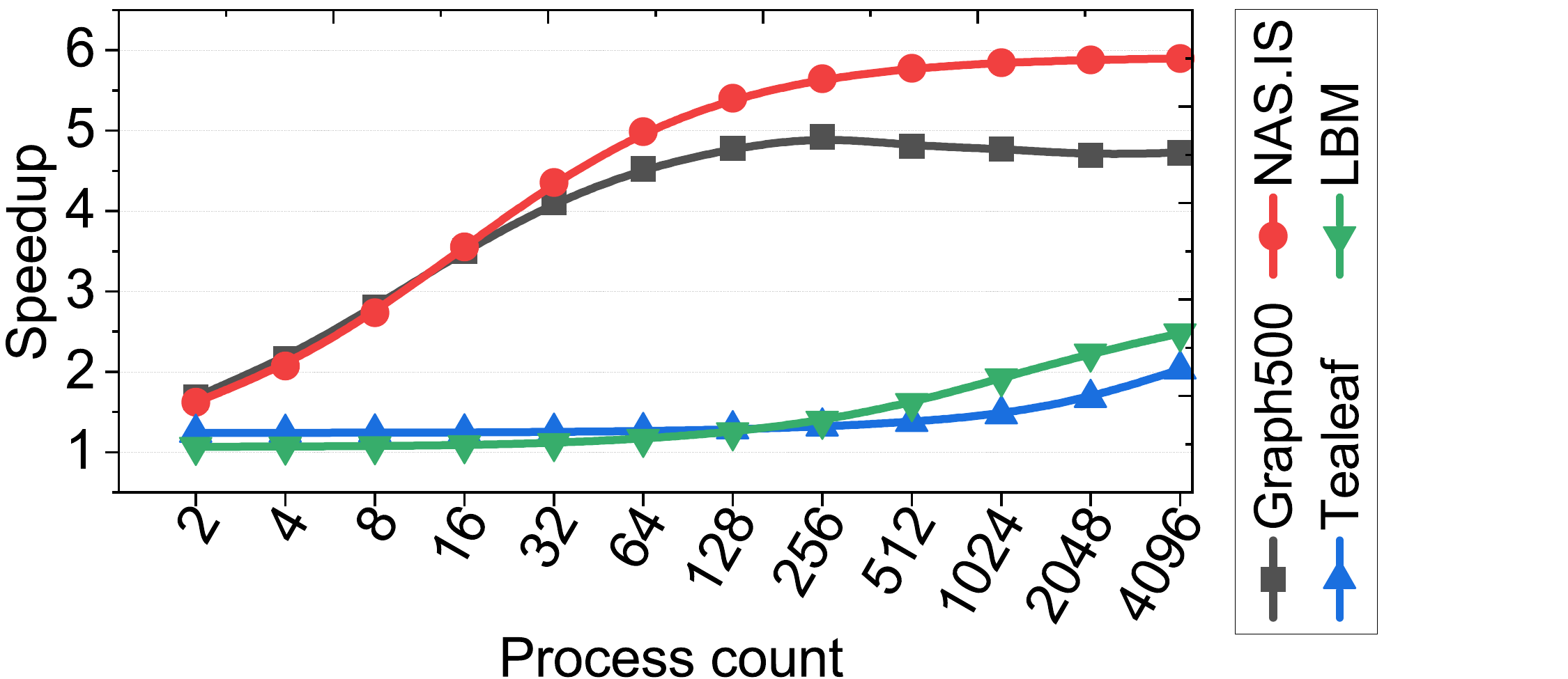}
    \caption{Strong scaling test.}
    \label{fig:eval_result2}
\end{figure}

Figure \ref{fig:eval_result2} presents the results of the strong scaling evaluation. Even for computation-intensive workloads such as Tealeaf and LBM, MPI-over-CXL achieved about 2.25$\times$ speedup over the baseline E2E execution-time at a 4,096 node scale. For communication-heavy benchmarks, Graph500 and NAS.IS, which primarily involve point-to-point and collective communication, respectively, the speedups were approximately 4.7$\times$ for Graph500 and 5.9$\times$ for NAS.IS. For communication-intensive workloads (e.g., Graph500, NAS.IS), communication time dominates total execution time, and applying MPI-over-CXL significantly reduces this overhead. This results in sharp speedup during early scaling stages until performance saturates as the benefits are maximized. On the other hand, for computation-intensive workloads (e.g., tealeaf, LBM), computation initially dominates, so communication improvements have little impact; however, the advantage of reducing communication latency gradually appears as the number of processes and communication overhead increases.

\section{Conclusion and Future Work }
In conclusion, MPI-over-CXL demonstrates significant advantages by reducing communication overhead through direct pointer-based interactions in coherent shared memory. The evaluation clearly illustrates performance improvements over traditional DSM systems, highlighting MPI-over-CXL's potential to significantly enhance HPC workload efficiency and scalability in future deployments. Future work involves evaluating MPI-over-CXL on larger computing clusters to confirm scalability and further performance gains. Additional optimizations to improve efficiency across diverse HPC applications will also be explored.

\nocite{*}
\balance
\bibliographystyle{IEEEtran}
\bibliography{references}

\begin{thebibliography}{10}
\providecommand{\url}[1]{#1}
\csname url@samestyle\endcsname
\providecommand{\newblock}{\relax}
\providecommand{\bibinfo}[2]{#2}
\providecommand{\BIBentrySTDinterwordspacing}{\spaceskip=0pt\relax}
\providecommand{\BIBentryALTinterwordstretchfactor}{4}
\providecommand{\BIBentryALTinterwordspacing}{\spaceskip=\fontdimen2\font plus
\BIBentryALTinterwordstretchfactor\fontdimen3\font minus
  \fontdimen4\font\relax}
\providecommand{\BIBforeignlanguage}[2]{{%
\expandafter\ifx\csname l@#1\endcsname\relax
\typeout{** WARNING: IEEEtran.bst: No hyphenation pattern has been}%
\typeout{** loaded for the language `#1'. Using the pattern for}%
\typeout{** the default language instead.}%
\else
\language=\csname l@#1\endcsname
\fi
#2}}
\providecommand{\BIBdecl}{\relax}
\BIBdecl

\bibitem{Padua2011}
D.~Padua, Ed., \emph{Encyclopedia of Parallel Computing}, 2011.

\bibitem{mpi1}
G.~Shipman, R.~Castain, R.~Graham, A.~Lumsdaine, G.~Bosilca, and B.~Barrett,
  ``{ Open MPI: A High-Performance, Heterogeneous MPI },'' in \emph{2013 IEEE
  International Conference on Cluster Computing (CLUSTER)}, 2006.

\bibitem{mpi2}
M.~G. Dosanjh, A.~Worley, D.~Schafer, P.~Soundararajan, S.~Ghafoor,
  A.~Skjellum, P.~V. Bangalore, and R.~E. Grant, ``Implementation and
  evaluation of mpi 4.0 partitioned communication libraries,'' \emph{Parallel
  Computing}, 2021.

\bibitem{vay2018warp}
J.-L. Vay, A.~Almgren, J.~Bell, L.~Ge, D.~Grote, M.~Hogan, O.~Kononenko,
  R.~Lehe, A.~Myers, C.~Ng \emph{et~al.}, ``{Warp-X: A new exascale computing
  platform for beam--plasma simulations},'' \emph{Nuclear Instruments and
  Methods in Physics Research Section A: Accelerators, Spectrometers, Detectors
  and Associated Equipment}, 2018.

\bibitem{spalart2016role}
P.~R. Spalart and V.~Venkatakrishnan, ``On the role and challenges of {CFD} in
  the aerospace industry,'' \emph{The Aeronautical Journal}, 2016.

\bibitem{Graph500}
{Graph500}, ``Graph 500 benchmarks,'' \url{https://graph500.org/}, 2010.

\bibitem{NASA.IS}
{NASA Advanced Supercomputing Division}, ``{NASA Parallel Benchmarks},''
  \url{https://www.nas.nasa.gov/software/npb.html}, 2024.

\bibitem{lauguna2019large}
I.~Laguna, R.~Marshall, K.~Mohror, M.~Ruefenacht, A.~Skjellum, and N.~Sultana,
  ``A large-scale study of mpi usage in open-source hpc applications,'' in
  \emph{Proceedings of the International Conference for High Performance
  Computing, Networking, Storage and Analysis (SC '19)}, 2019.

\bibitem{mpiapp}
S.~Ghosh, M.~Halappanavar, A.~Kalyanaraman, A.~Khan, and A.~H. Gebremedhin,
  ``{Exploring MPI Communication Models for Graph Applications Using Graph
  Matching as a Case Study},'' in \emph{2019 IEEE International Parallel and
  Distributed Processing Symposium (IPDPS'19)}, 2019.

\bibitem{haloapp}
L.~Spies, A.~Bienz, D.~Moulton, L.~Olson, and A.~Reisner, ``Tausch: A halo
  exchange library for large heterogeneous computing systems using mpi, opencl,
  and cuda,'' \emph{Parallel Computing}, 2022.

\bibitem{zambre2019breaking}
R.~Zambre, M.~Grodowitz, A.~Chandramowlishwaran, and P.~Shamis, ``{Breaking
  band: A breakdown of high-performance communication},'' in \emph{Proceedings
  of the 48th International Conference on Parallel Processing (ICPP'19)}, 2019.

\bibitem{mpioverhead}
T.~Hoefler, J.~Dinan, R.~Thakur, B.~Barrett, P.~Balaji, W.~Gropp, and
  K.~Underwood, ``Remote memory access programming in mpi-3,'' \emph{ACM
  Transactions on Parallel Computing (TOPC)}, 2015.

\bibitem{mpioverhead2}
M.~P{\'e}rache, P.~Carribault, and H.~Jourdren, ``Mpc-mpi: An mpi
  implementation reducing the overall memory consumption,'' in \emph{European
  Parallel Virtual Machine/Message Passing Interface Users' Group Meeting},
  2009.

\bibitem{mpioverhead4}
J.~Liu, J.~Wu, S.~P. Kini, P.~Wyckoff, and D.~K. Panda, ``{High performance
  RDMA-based MPI implementation over InfiniBand},'' in \emph{Proceedings of the
  17th Annual International Conference on Supercomputing (ICS'03)}, 2003.

\bibitem{cxl3}
{CXL Consortium}, ``{CXL} 3.2 specification,''
  \url{https://computeexpresslink.org/cxl-specification/}, 2024.

\bibitem{memsharing}
S.~Jain, N.~Yeleswarapu, H.~A. Maruf, and R.~Gupta, ``{Memory Sharing with CXL:
  Hardware and Software Design Approaches},'' https://arxiv.org/abs/2404.03245,
  2024.

\bibitem{memsharing1}
D.~Gouk, M.~Kwon, H.~Bae, S.~Lee, and M.~Jung, ``{Memory Pooling With CXL},''
  \emph{IEEE Micro}, 2023.

\bibitem{memsharing2}
M.~Ha, J.~Ryu, J.~Choi, K.~Ko, S.~Kim, S.~Hyun, D.~Moon, B.~Koh, H.~Lee, M.~Kim
  \emph{et~al.}, ``Dynamic capacity service for improving cxl pooled memory
  efficiency,'' \emph{IEEE Micro}, 2023.

\bibitem{huang2025txcocket}
T.~Huang, Y.~Liang, S.~Yu, and K.~Chen, ``Txcocket: an innovative solution for
  efficient cross-node data transmission enabled by cxl-based shared memory,''
  \emph{CCF Transactions on High Performance Computing}, 2025.

\bibitem{zhang2023partial}
M.~Zhang, T.~Ma, J.~Hua, Z.~Liu, K.~Chen, N.~Ding, F.~Du, J.~Jiang, T.~Ma, and
  Y.~Wu, ``Partial failure resilient memory management system for (cxl-based)
  distributed shared memory,'' in \emph{Proceedings of the 29th Symposium on
  Operating Systems Principles}, 2023.

\bibitem{shmem}
T.~Hoefler, J.~Dinan, D.~Buntinas, P.~Balaji, B.~W. Barrett, R.~Brightwell,
  W.~Gropp, V.~Kale, and R.~Thakur, ``{Leveraging MPI's One-Sided Communication
  Interface for Shared-Memory Programming},'' in \emph{Recent Advances in the
  Message Passing Interface (EuroMPI'12)}, 2012.

\bibitem{openshmem1}
J.~R. Hammond, S.~Ghosh, and B.~M. Chapman, ``{Implementing OpenSHMEM Using
  MPI-3 One-Sided Communication},'' in \emph{OpenSHMEM and Related
  Technologies. Experiences, Implementations, and Technologies (OpenSHMEM'14)},
  2014.

\bibitem{sur2006rdma}
S.~Sur, H.-W. Jin, L.~Chai, and D.~K. Panda, ``Rdma read based rendezvous
  protocol for mpi over infiniband: design alternatives and benefits,'' in
  \emph{Proceedings of the eleventh ACM SIGPLAN symposium on Principles and
  practice of parallel programming}, 2006, pp. 32--39.

\bibitem{wu2004high}
J.~Wu, P.~Wyckoff, and D.~Panda, ``High performance implementation of mpi
  derived datatype communication over infiniband,'' in \emph{18th International
  Parallel and Distributed Processing Symposium, 2004. Proceedings.}, 2004.

\bibitem{traff2023uniform}
J.~L. Tr{\"a}ff, S.~Hunold, I.~Vardas, and N.~M. Funk, ``Uniform algorithms for
  reduce-scatter and (most) other collectives for mpi,'' in \emph{2023 IEEE
  International Conference on Cluster Computing (CLUSTER)}, 2023.

\bibitem{gainaru2016using}
A.~Gainaru, R.~L. Graham, A.~Polyakov, and G.~Shainer, ``Using infiniband
  hardware gather-scatter capabilities to optimize mpi all-to-all,'' in
  \emph{Proceedings of the 23rd European MPI Users' Group Meeting}, 2016.

\bibitem{keppens2023mpi}
R.~Keppens, B.~P. Braileanu, Y.~Zhou, W.~Ruan, C.~Xia, Y.~Guo, N.~Claes, and
  F.~Bacchini, ``Mpi-amrvac 3.0: Updates to an open-source simulation
  framework,'' \emph{Astronomy \& Astrophysics}, 2023.

\bibitem{ly2023alisim}
N.~Ly-Trong, G.~M. Barca, and B.~Q. Minh, ``Alisim-hpc: parallel sequence
  simulator for phylogenetics,'' \emph{Bioinformatics}, 2023.

\bibitem{memorycopyredun}
J.~Peng, J.~Fang, J.~Liu, M.~Xie, Y.~Dai, B.~Yang, S.~Li, and Z.~Wang,
  ``{Optimizing MPI Collectives on Shared Memory Multi-Cores},'' in
  \emph{Proceedings of the International Conference for High Performance
  Computing, Networking, Storage and Analysis (SC'23)}, 2023.

\bibitem{chen2023mpi}
C.-C. Chen, K.~Shafie~Khorassani, P.~Kousha, Q.~Zhou, J.~Yao, H.~Subramoni, and
  D.~K. Panda, ``Mpi-xccl: A portable mpi library over collective communication
  libraries for various accelerators,'' in \emph{Proceedings of the SC'23
  Workshops of the International Conference on High Performance Computing,
  Network, Storage, and Analysis}, 2023.

\bibitem{mpi5}
{MPI Forum}, ``{MPI}: A message-passing interface standard, version 5.0,''
  \url{https://www.mpi-forum.org/docs/mpi-5.0/mpi50-report.pdf}, 2021.

\bibitem{multiplenode}
B.~Ramesh, K.~K. Suresh, N.~Sarkauskas, M.~Bayatpour, J.~M. Hashmi,
  H.~Subramoni, and D.~K. Panda, ``{Scalable MPI Collectives using SHARP: Large
  Scale Performance Evaluation on the TACC Frontera System},'' in \emph{2020
  Workshop on Exascale MPI (ExaMPI'20)}, 2020.

\bibitem{balaji2010}
P.~Balaji, A.~Chan, W.~Gropp, R.~Thakur, and E.~Lusk, ``The importance of
  non-data-communication overheads in {MPI},'' \emph{The International Journal
  of High Performance Computing Applications}, 2010.

\bibitem{bierbaum2022towards}
J.~Bierbaum, M.~Planeta, and H.~H{\"a}rtig, ``Towards efficient
  oversubscription: on the cost and benefit of event-based communication in
  mpi,'' in \emph{2022 IEEE/ACM International Workshop on Runtime and Operating
  Systems for Supercomputers (ROSS)}, 2022.

\bibitem{mpinodecomlexity}
G.~Bosilca, T.~Herault, A.~Rezmerita, and J.~Dongarra, ``{On Scalability for
  MPI Runtime Systems},'' in \emph{2011 IEEE International Conference on
  Cluster Computing (CLUSTER'11)}, 2011.

\bibitem{mpimanyprocessor}
P.~Balaji, D.~Buntinas, D.~Goodell, W.~Gropp, S.~Kumar, E.~Lusk, R.~Thakur, and
  J.~L. Tr{\"a}ff, ``{MPI on a Million Processors},'' in \emph{Recent Advances
  in Parallel Virtual Machine and Message Passing Interface (PVM/MPI'09)},
  2009.

\bibitem{hofinger2017modelling}
S.~H{\"o}finger and E.~Haunschmid, ``Modelling parallel overhead from simple
  run-time records,'' \emph{The Journal of Supercomputing}, 2017.

\bibitem{pcie}
{PCI-SIG}, ``{{PCI} Express Base Specification Revision 6.0},'' \url{
  https://pcisig.com/specifications}, 2022.

\bibitem{gouk2022direct}
D.~Gouk, S.~Lee, M.~Kwon, and M.~Jung, ``Direct access, {High-Performance}
  memory disaggregation with {DirectCXL},'' in \emph{2022 USENIX Annual
  Technical Conference (USENIX ATC 22)}, 2022.

\bibitem{cxl1}
{CXL Consortium}, ``{CXL} 1.1 specification,''
  \url{https://computeexpresslink.org/past-cxl-specifications/}, 2019.

\bibitem{cxl2}
{CXL Consortium}, ``{CXL} 2.0 specification,''
  \url{https://computeexpresslink.org/past-cxl-specifications/}, 2020.

\bibitem{santhanaraman2005designing}
G.~Santhanaraman, J.~Wu, W.~Huang, and D.~K. Panda, ``Designing zero-copy
  message passing interface derived datatype communication over infiniband:
  Alternative approaches and performance evaluation,'' \emph{The International
  Journal of High Performance Computing Applications}, 2005.

\bibitem{hashmi2020falcon}
J.~M. Hashmi, C.-H. Chu, S.~Chakraborty, M.~Bayatpour, H.~Subramoni, and D.~K.
  Panda, ``Falcon-x: Zero-copy mpi derived datatype processing on modern cpu
  and gpu architectures,'' \emph{Journal of Parallel and Distributed
  Computing}, 2020.

\bibitem{ma2021thinking}
T.~Ma, K.~Chen, S.~Ma, Z.~Song, and Y.~Wu, ``Thinking more about rdma memory
  semantics,'' in \emph{2021 IEEE International Conference on Cluster Computing
  (CLUSTER)}, 2021.

\bibitem{kong2023understanding}
X.~Kong, J.~Chen, W.~Bai, Y.~Xu, M.~Elhaddad, S.~Raindel, J.~Padhye, A.~R.
  Lebeck, and D.~Zhuo, ``Understanding rdma microarchitecture resources for
  performance isolation,'' in \emph{20th USENIX Symposium on Networked Systems
  Design and Implementation (NSDI 23)}, 2023.

\bibitem{opt1}
G.~Congiu and P.~Balaji, ``{Evaluating the Impact of High-Bandwidth Memory on
  MPI Communications},'' in \emph{2018 IEEE 4th International Conference on
  Computer and Communications (ICCC'18)}, 2018.

\bibitem{opt2}
Q.~Zhou, P.~Kousha, Q.~Anthony, K.~Shafie~Khorassani, A.~Shafi, H.~Subramoni,
  and D.~K. Panda, ``{Accelerating MPI All-to-All Communication with Online
  Compression on Modern GPU Clusters},'' in \emph{High Performance Computing
  (HiPC'22)}, 2022.

\bibitem{tealeaf}
S.~McIntosh-Smith, M.~Martineau, T.~Deakin, G.~Pawelczak, W.~Gaudin,
  P.~Garrett, W.~Liu, R.~Smedley-Stevenson, and D.~Beckingsale, ``Tealeaf: A
  mini-application to enable design-space explorations for iterative sparse
  linear solvers,'' in \emph{2017 IEEE International Conference on Cluster
  Computing (CLUSTER'17)}, 2017.

\bibitem{LBM}
{Lattice Boltzmann Method}, ``{Lattice Boltzmann Method Benchmarks},''
  \url{https://www.spec.org/cpu2017/Docs/benchmarks/619.lbm_s.html}, 2017.

\bibitem{cai2018efficient}
Q.~Cai, W.~Guo, H.~Zhang, D.~Agrawal, G.~Chen, B.~C. Ooi, K.-L. Tan, Y.~M. Teo,
  and S.~Wang, ``{Efficient distributed memory management with RDMA and
  caching},'' \emph{Proceedings of the VLDB Endowment}, 2018.

\bibitem{spechpc}
A.~Afzal, G.~Hager, and G.~Wellein, ``{SPEChpc 2021 Benchmarks on Ice Lake and
  Sapphire Rapids Infiniband Clusters: A Performance and Energy Case Study},''
  in \emph{Proceedings of the SC '23 Workshops of the International Conference
  on High Performance Computing, Network, Storage, and Analysis (SC-W'23)},
  2023.

\bibitem{chester2021stressbench}
D.~G. Chester, T.~Groves, S.~D. Hammond, T.~Law, S.~A. Wright,
  R.~Smedley-Stevenson, S.~A. Fahmy, G.~R. Mudalidge, and S.~A. Jarvis,
  ``Stressbench: a configurable full system network and i/o benchmark
  framework,'' in \emph{2021 IEEE High Performance Extreme Computing Conference
  (HPEC'21)}, 2021.

\bibitem{heattransfer}
Å.~Łach and D.~Svyetlichnyy, ``{Advances in Numerical Modeling for Heat
  Transfer and Thermal Management: A Review of Computational Approaches and
  Environmental Impacts},'' \emph{Energies}, 2025.

\end{thebibliography}

\end{document}